\pdfoutput=1 
\RequirePackage{ifpdf}
\documentclass{JINST}

\title{Characterisation of Glass Electrodes and RPC Detectors for $INO-ICAL$ Experiment}

\author{Md. Naimuddin,
Daljeet Kaur,
Purnendu Kumar, Ankit Gaur, Praveen Kumar, 
Md. Hasbuddin, Swati Mishra and Ashok Kumar\\
Department of Physics \& Astrophysics, University of Delhi,\\
  Delhi - 110 007, India\\
E-mail: \email{nayeem@cern.ch}}

\abstract{India-based Neutrino Observatory (INO) is a planned neutrino experiment to be 
build up in southern part of India.The INO observatory will host a 51 kton Iron Calorimeter (ICAL) detector to detect atmospheric neutrinos. Resistive Plate Chamber (RPC) has been chosen as the active detector element for the ICAL experiment. The ICAL experiment will consist of about 28,000 RPC's of dimension $2~m\times 2~m$, divided into three modules. The experiment is planned to take data at least for 20 years from its start date. Due to the large number of RPC needed for ICAL experiment and the long lifetime of the experiment, it is necessary to carry out 
detailed $R\&D$ to optimise each and every parameter of the detector performance.  We report on the performance studies carried out on the RPC's made with these electrodes, and finally compare the detector performance with that of the material properties to optimise the detector parameters.}


\begin{document}

\section{Introduction}\label{sec:intro}
The India-based Neutrino Observatory (INO)~\cite{ino} is a mega science project aimed at building a large underground laboratory to study the atmospheric neutrinos. The observatory will be a multi-experiment facility with Iron CALorimeter (ICAL) detector as one of the experiments. The entire ICAL detector will sit under a rock cover of approximately 1 km and will be magnetized to detect the charge of muons produced from neutrinos. Neutrinos are fundamental particles belonging to the lepton family in the standard model of particle physics and are assumed massless. However, recent experiments indicate that neutrinos oscillate among their flavors and have finite but tiny masses that are yet to be measured. Determination of neutrino masses and oscillation parameters is one of the most important open problems in physics today. INO-ICAL experiment with active resistive plate chamber detector will be designed to address these problems by detecting the passage of charged particles mainly muons produced due to neutrino interactions with the ICAL detector.

\section{Resistive Plate Chambers}\label{sec:rpc}
Resistive Plate Chamber (RPC) detectors~\cite{Santonico1981, Santonico1988} are the 
gaseous detectors pioneered during 1980s. Since then it has found its usage in various 
high energy, nuclear and astroparticle physics experiments as well as applications in  cargo imaging, medical diagnostic imaging and other fields. 
The RPC detectors are known for its long time stability and large coverage area for detection. These detectors also produce very large pulse heights, in the range of 2-5 mV without any external amplification, when operating in avalanche mode.  In addition, RPC has excellent time resolution (few $ns$), and good detection efficiency for minimum ionising particles. RPC can also be used to build a fast, efficient and robust triggering system by making use of its excellent time resolution, good efficiency and high granularity~\cite{Bencze}. The RPC's are also very simple and low cost detectors. However, it is very important to carefully choose the various design parameters like the electrode material, gas composition, environmental factors, etc. to fully exploit all the advantages of the RPC detectors. 

Electrode material plays crucial 
role in the functioning of the RPC detectors, so it is very important to select it carefully. The electrode material should have high resistivity and 
 high surface smoothness to avoid localisation of excess charge, and to prevent 
alternating leakage path for post streamer recovery. Bakelite and glass are the most sutiable and commonly used electrode materials for the RPC detectors. We are exploring both of these materials to be used for the INO RPC detectors. In this paper, we provide results of our studies performed for the glass electrodes only. The details of our bakelite studies can be found elsewhere~\cite{ourbakelite}. 

We selected three types of glass viz., Saint Gobain, Asahi and Modi which are available in the local market for our $R\& D$. In order to determine the electrical properties of the glasses 
we performed the bulk resistivity and surface resistivity measurements. 
For the purpose of determining 
the smoothness of the surface we performed Scanning Electron 
Microscopy (SEM) and Atomic Force Microscopy (AFM) measurements. We also 
performed the X-Ray Diffractometry (XRD) measurements to determine the crystal 
structure of the glass samples. Ultra Violet (UV) transmittance test was 
performed to find out the the 
reflectance and transmittance capabilities of these glasses. 

Bulk resistivity of the Saint Gobain glass is found to be maximum, which is of the order $8 \times 10^{12}$ $\Omega$-cm. Modi and Asahi are having similar bulk resistivity and is about 3 times smaller than Saint Gobain. The surface resistivity for Asahi and Modi are similar and is of the order of 
$2\times10^{11}\Omega/\Box$ on average. The Saint Gobain surface resistivity is slightly smaller and is of the order of $1.6\times 10^{11} \Omega/\Box$ on average.

The rough inner surface of the electrodes can cause variations in the electric 
field inside the RPC, so it is very important to have the inner surface as smooth as 
possible for the efficient operation of the RPC. Asahi is found to have the smoothest surface followed by Saint Gobain and Modi respectively. The XRD measurements of the glass samples shows amorphous structure for all the samples. We performed transmittance test on Perkin Elmer Lamda 3B spectrophotometer between Lambda range of 0 - 1200 $nm$ and found Saint Gobain to be most reflective while Modi the least with Asahi in between.

\section{RPC Characterisation}\label{sec:char}
Using the three types of glasses of thickness 3~$mm$, we fabricated several prototypes of size $30~cm\times 30~cm$ RPC's. Glass plate electrodes were coated with a conductive layer of graphite on the outer surface. For maintaining a uniform gas gap, polycarbonate button spacer of thickness 2~$mm$ were glued between the two electrode plates. The gas gaps were sealed from all sides using side spacers made up of polycarbonate. A gas inlet and 
a gas outlet nozzles were glued on the diagonally opposite corners of the chamber. Honey comb panels with strips of width 2.8 $cm$ were used as pick up panels. Sealed RPC chambers were then checked for any gas leakage via a leak test using manometer technique.

The fabricated RPC's were characterised for efficiency, leakage current and noise rate under different operating conditions. We 
characterise the RPC's for various gas composition, environmental temperature, atmospheric humidity and 
threshold to obtain the optimum parameters to maximise the detector performance. A muon telescope consisting of three scintillator detectors 
connected with the NIM/VME Data Acquisition system (DAQ) were put in place. The RPC detector, which was to be characterised, was squeezed in between the scintillator detectors and readout by the DAQ system.

\subsection{Variation of Gas Mixture Composition}
\begin{figure}[H]
\begin{minipage}{\linewidth}
      \begin{minipage}{0.3\linewidth}
          \includegraphics[width=5cm,height=5cm]{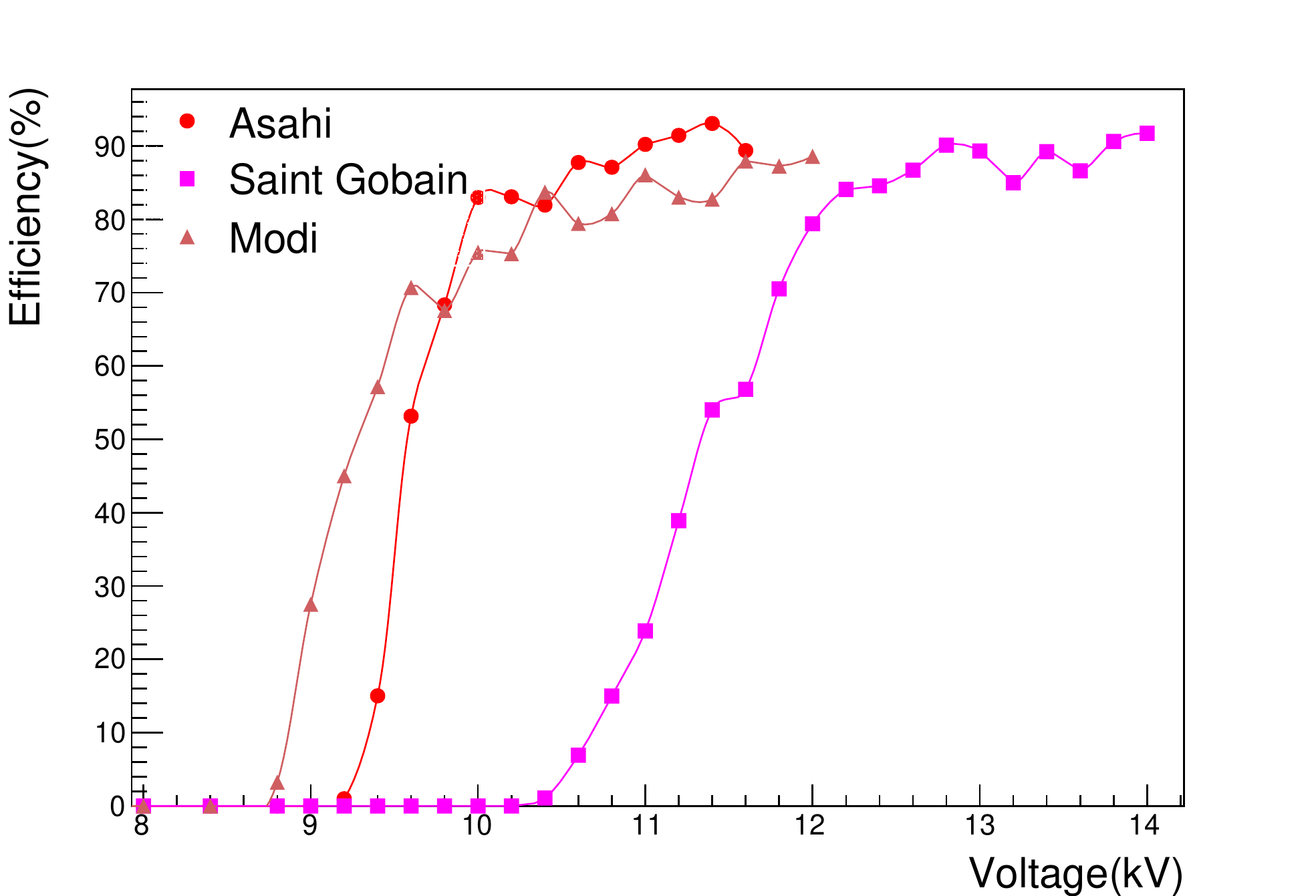}
               \end{minipage}
      \begin{minipage}{0.3\linewidth}
            \includegraphics[width=5cm,height=5cm]{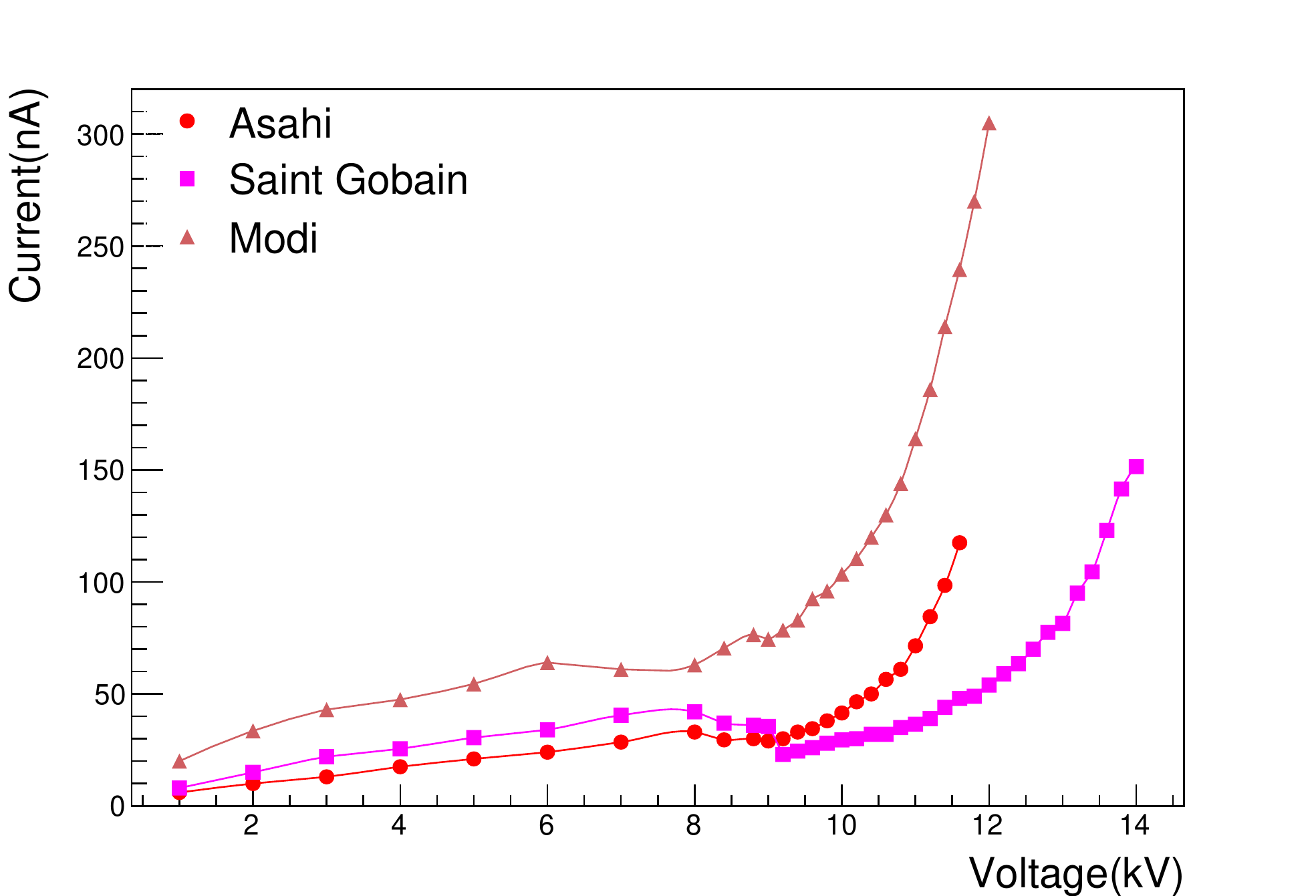}
             \end{minipage}
          \begin{minipage}{0.3\linewidth}
                 \includegraphics[width=5cm,height=5cm]{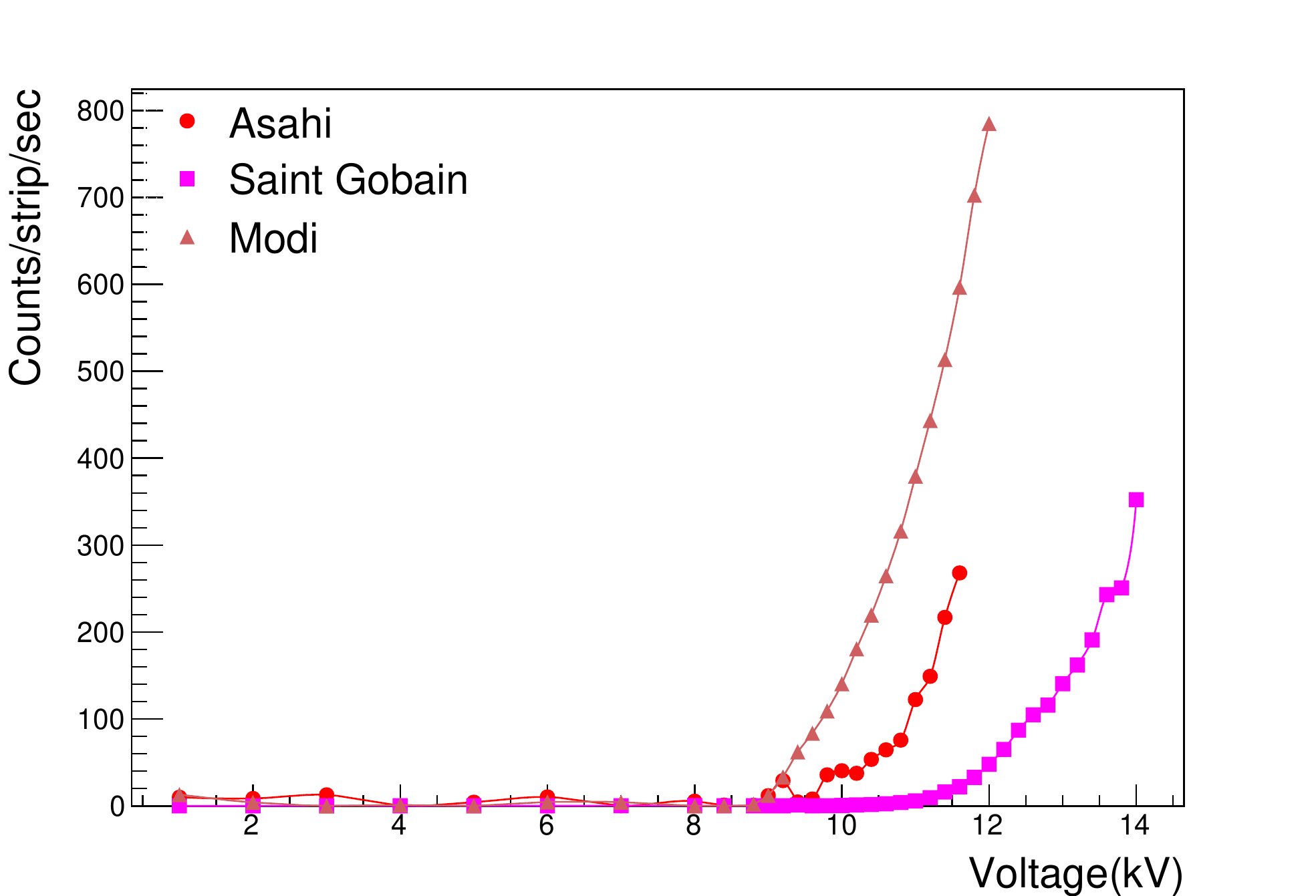}
        \end{minipage}
          \end{minipage}
          \caption{Efficiency, Leakage Current and Noise rates for the $R134a$ (95.0\%), $C_{4}H_{10}$ (4.5\%), $SF_{6}$ (0.5\%) gas mixture.}
\label{fig:GasMix1}
 \end{figure}

\begin{figure}[H]
\begin{minipage}{\linewidth}
      \centering
      \begin{minipage}{0.3\linewidth}
          \includegraphics[width=5cm,height=5cm]{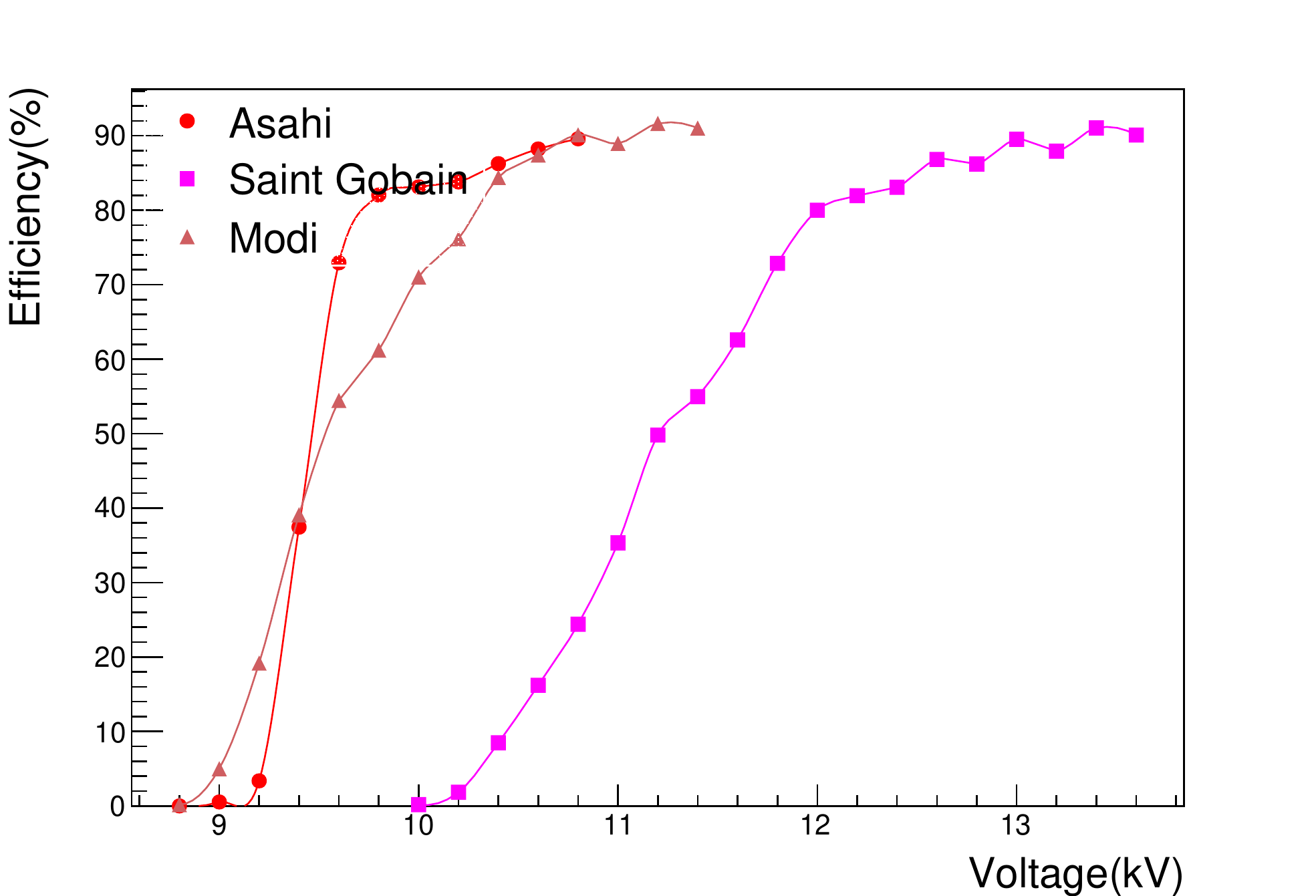}
               \end{minipage}
      \begin{minipage}{0.3\linewidth}
            \includegraphics[width=5cm,height=5cm]{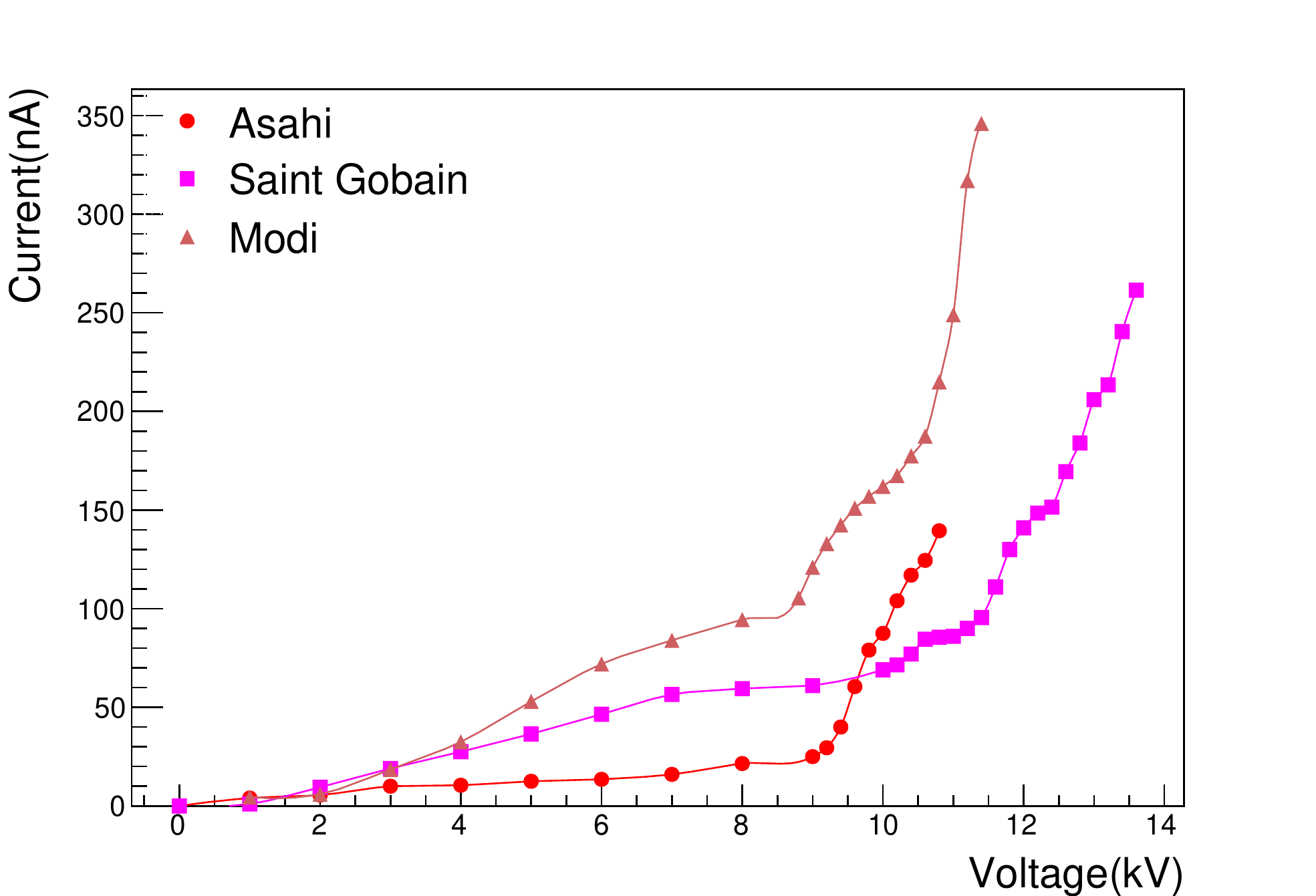}
             \end{minipage}
\begin{minipage}{0.3\linewidth}
                 \includegraphics[width=5cm,height=5cm]{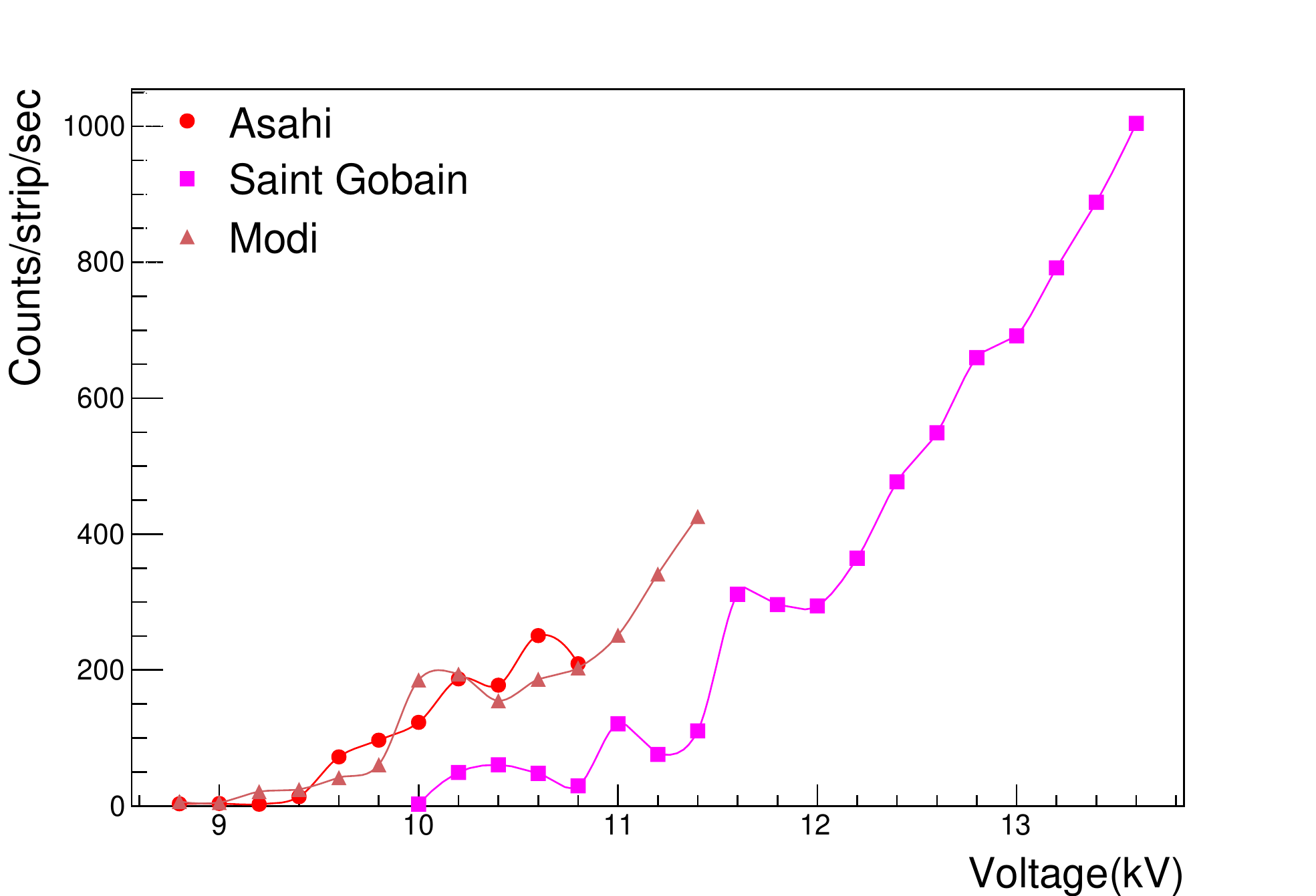}
        \end{minipage}
          \end{minipage}
          \caption{Efficiency and Leakage Current for the $R134a$ (67.7\%), $C_{4}H_{10}$ (32.0\%), $SF_{6}$ (0.3\%) Gas Mixture.}
\label{fig:GasMix2}
 \end{figure}

\begin{figure}[H]
\begin{minipage}{\linewidth}
      \centering
      \begin{minipage}{0.3\linewidth}
          \includegraphics[width=5cm,height=5cm]{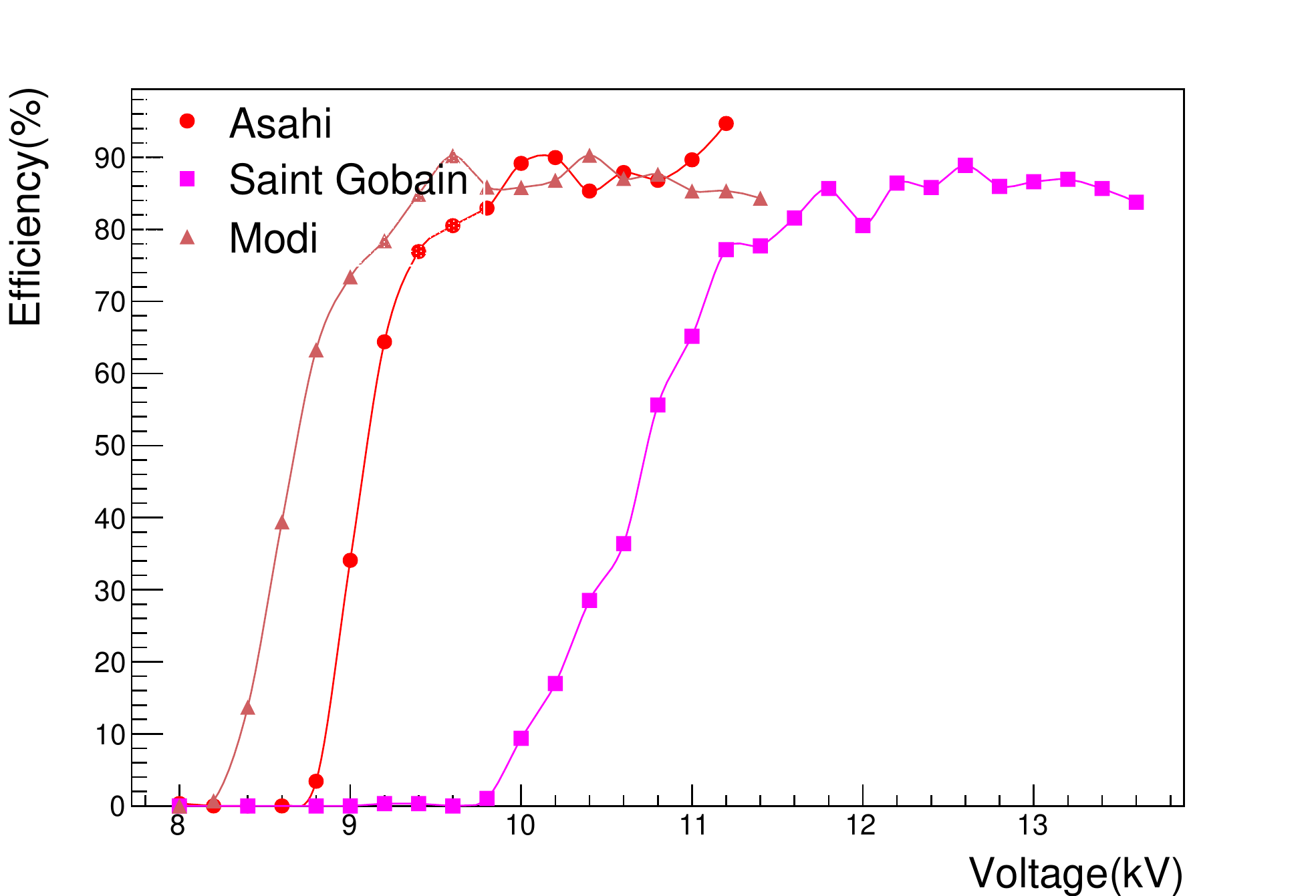}
               \end{minipage}
      \begin{minipage}{0.3\linewidth}
            \includegraphics[width=5cm,height=5cm]{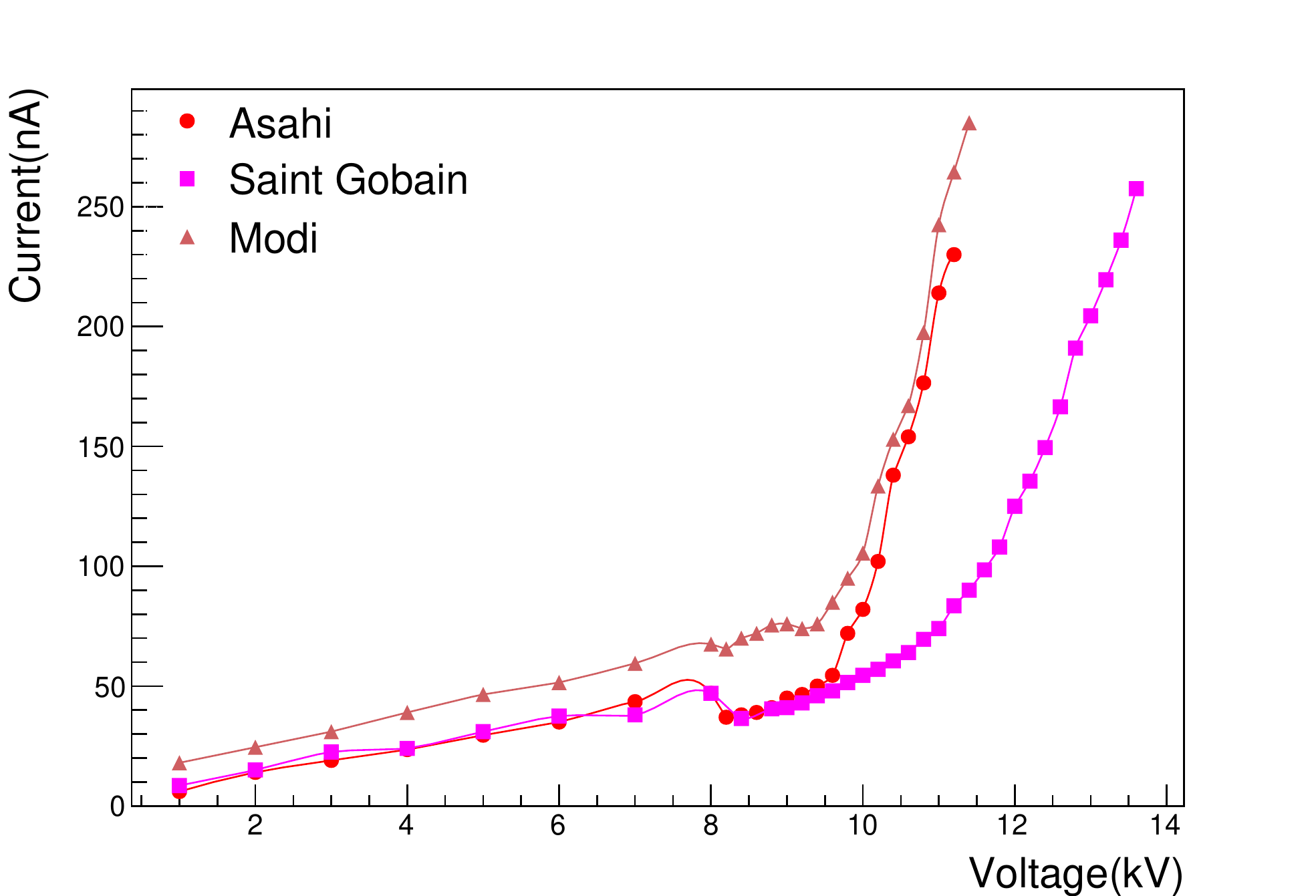}
             \end{minipage}
\begin{minipage}{0.3\linewidth}
                 \includegraphics[width=5cm,height=5cm]{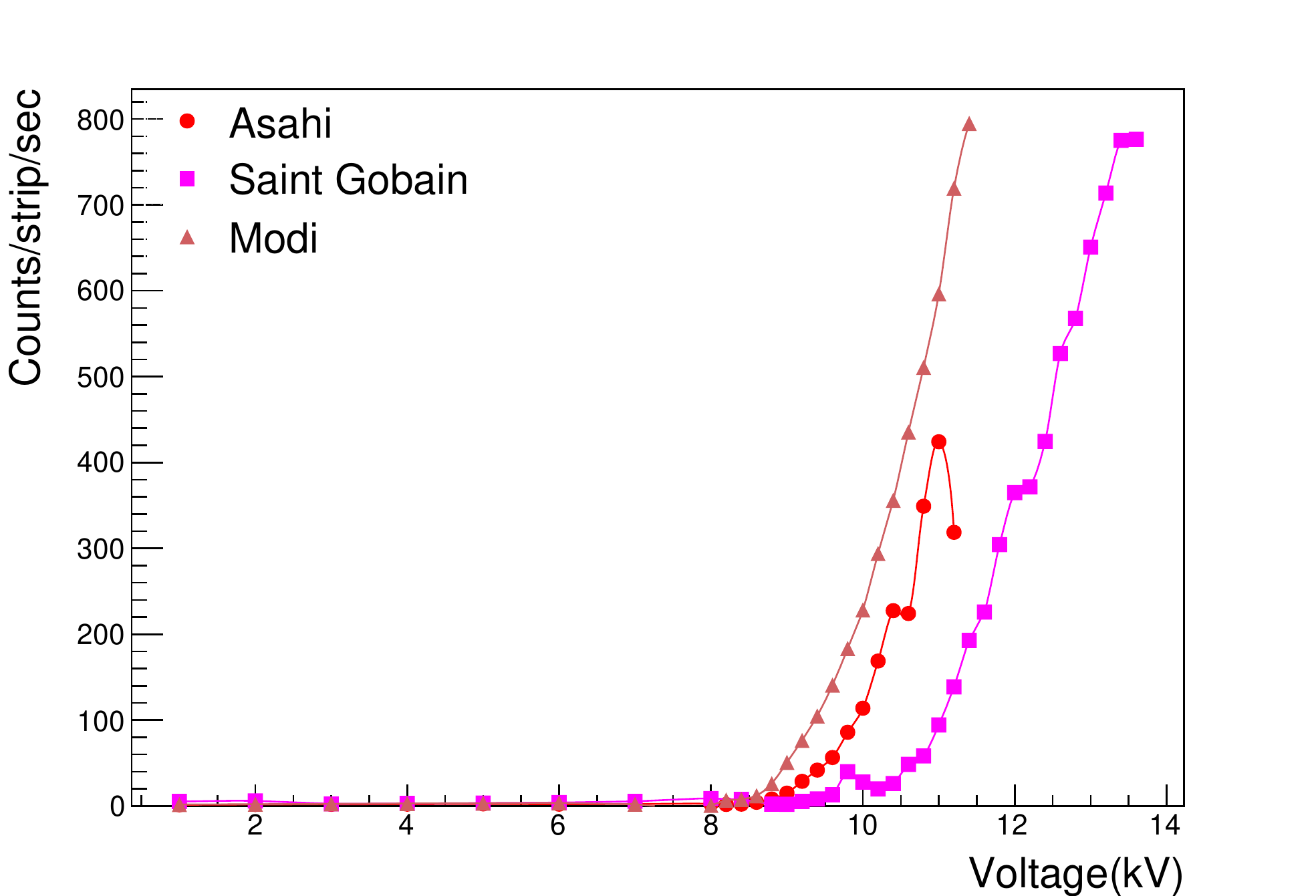}
        \end{minipage}
          \end{minipage}
          \caption{Efficiency and Leakage Current for $R134a$ (95.0\%), $C_{4}H_{10}$ (5.0\%), $SF_{6}$ (0.0\%) Gas Mixture.}
\label{fig:GasMix3}
 \end{figure}

The RPC's were tested in avalanche mode with a gas mixture of $R134a$ (95.0\%), $C_{4}H_{10}$ (4.5\%), $SF_{6}$ (0.5\%) with a flow rate of 10 SCCM. This mixture is being currently used to  operate the INO-ICAL RPC's prototypes~\cite{Pramana}. However, it is important to determine whether 
this composition is the optimum or not. For this purpose we varied the compositions 
of these gases and measured efficiency, noise rate and leakage current of the RPC's. Fig.~\ref{fig:GasMix1} shows the efficiency, leakage current and noise rate  
for the $R134a$ (95.0\%), $C_{4}H_{10}$ (4.5\%), $SF_{6}$ (0.5\%) gas composition. Fig.~\ref{fig:GasMix2} and  Fig.~\ref{fig:GasMix3} shows the efficiency and leakage current for the gas composition of $R134a$ (67.7\%), $C_{4}H_{10}$ (32.0\%), $SF_{6}$ (0.3\%) and $R134a$ (95.0\%), $C_{4}H_{10}$ (5.0\%), $SF_{6}$ (0.0\%) respectively. From these figures it can be seen that the efficiencies 
for all these gas compositions are similar for all types of glasses. In all the chambers the plaeteau is about 90\%. The inefficiency is due to misalignment of the triggering paddles. With the proper alignment of triggering paddles the efficiency increases upto 97\%. Noise rate as expected is maximum in the absence of $SF_{6}$~\cite{gas}. Also, the noise rate is maximum for Modi and 
minimum for Asahi. Leakage current is much higher for Modi compared to Asahi and Saint 
Gobain, which have reasonable leakage current.

\subsection{Variation of environmental Temperature and Humidity}

\begin{figure}[H]
\begin{minipage}{\linewidth}
        \begin{minipage}{0.3\linewidth}
          \includegraphics[width=5cm,height=5cm]{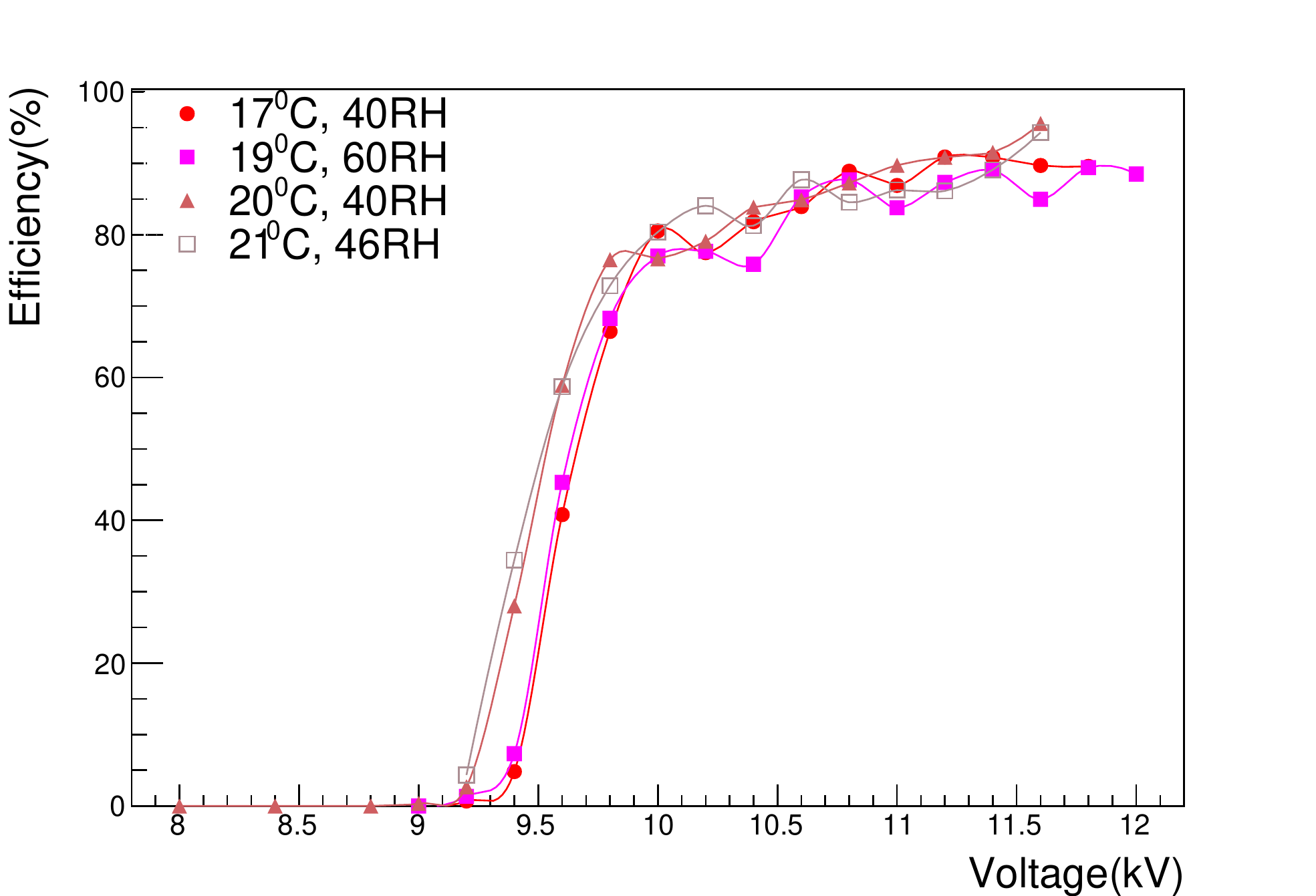}
               \end{minipage}
      \begin{minipage}{0.3\linewidth}
            \includegraphics[width=5cm,height=5cm]{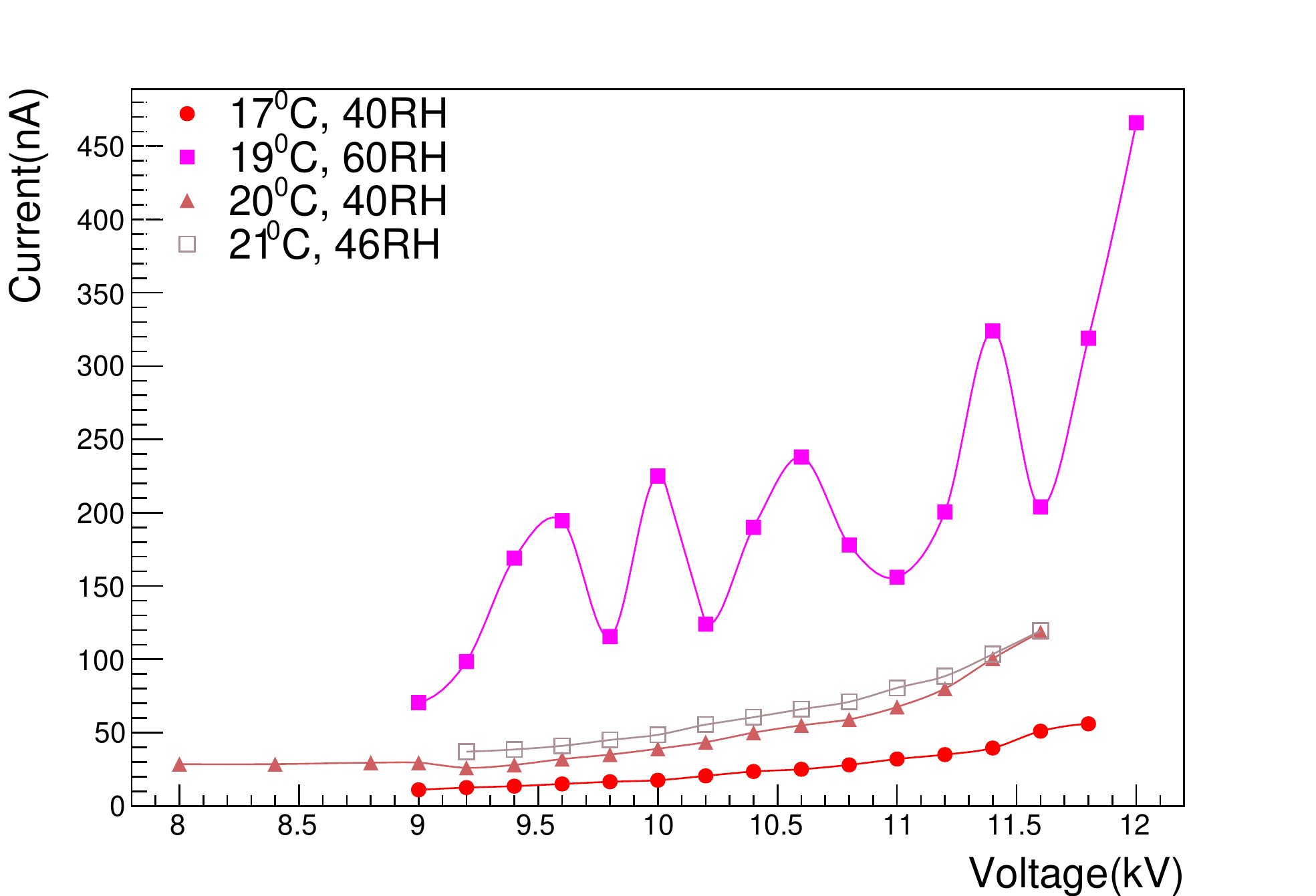}
             \end{minipage}
          \begin{minipage}{0.3\linewidth}
            \includegraphics[width=5cm,height=5cm]{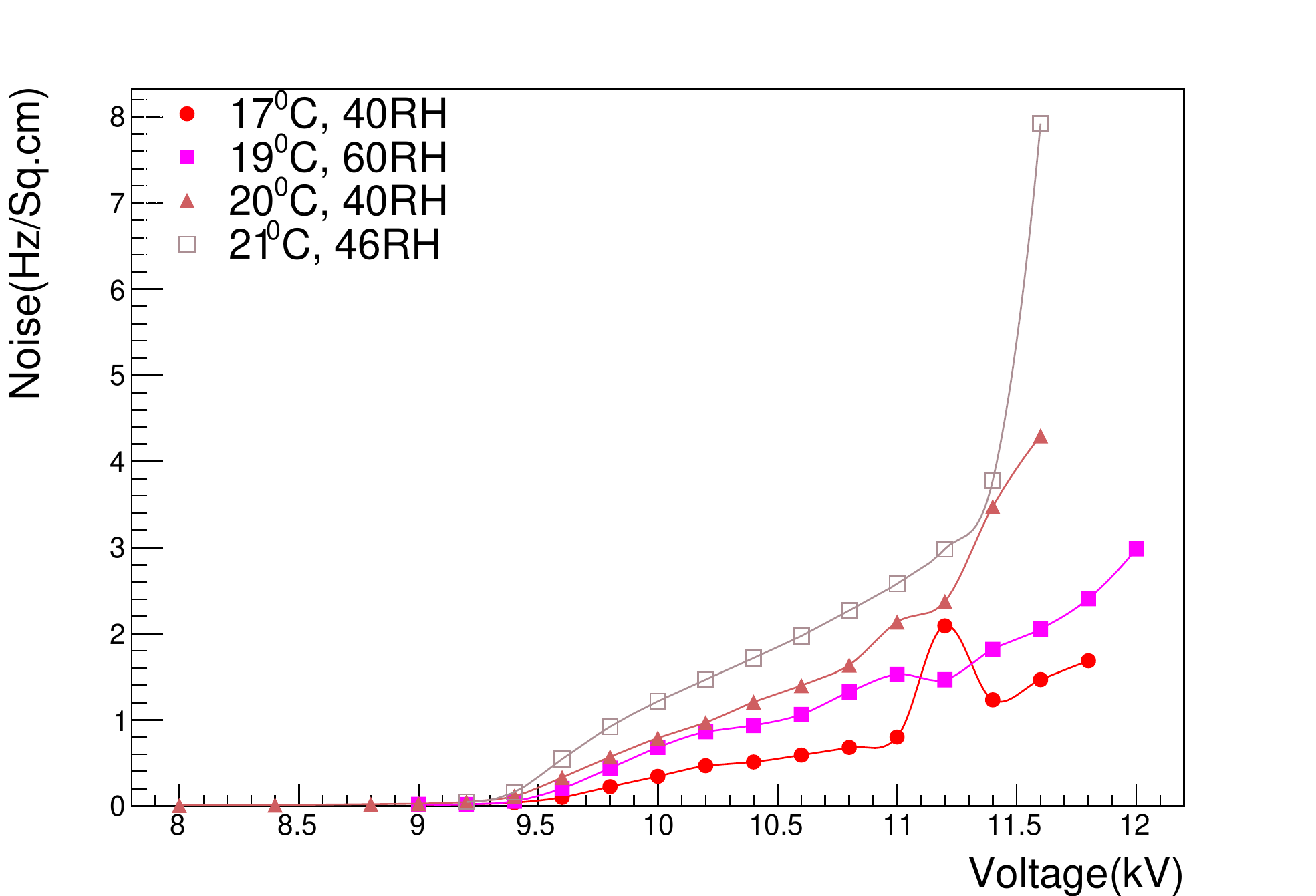}
        \end{minipage}
          \end{minipage}
          \caption{Efficiency, Leakage Current and Noise rates as a function of temperature and humidity for $R134a$ (95.0\%), $C_{4}H_{10}$ (4.5\%), $SF_{6}$ (0.5\%) gas mixture. }
\label{fig:Temp}
 \end{figure}

Operational conditions like environment temperature and moisture also affects the performance 
of the RPC detectors, so it is very important to find out the suitable environmental 
conditions to operate the RPC in order to optimise the performance. We varied the environmental 
temperature and humidity to different values and measured the efficiency, noise rate and 
leakage current for the RPCs. Fig.~\ref{fig:Temp} shows the effect of temperature and humidity variation on the efficiency, 
current and noise rate for the RPC made up of Asahi glass. Saint Gobain and Modi RPC's shows similar behaviour.

\subsection{Variation of Thresholds} 

\begin{figure}[H]

\begin{minipage}{\linewidth}
      \centering
      \begin{minipage}{0.3\linewidth}
          \includegraphics[width=5cm,height=5cm]{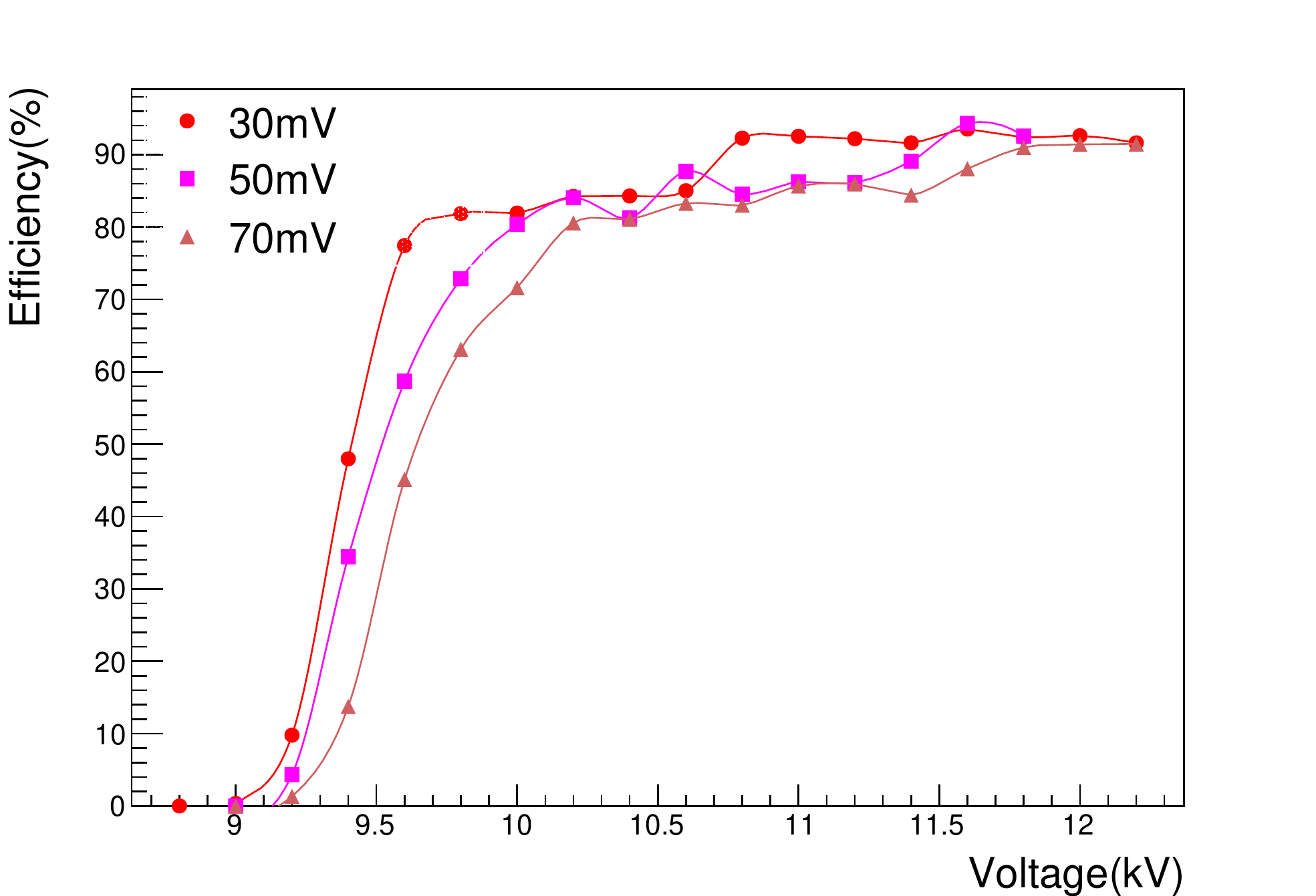}
               \end{minipage}
      \begin{minipage}{0.3\linewidth}
            \includegraphics[width=5cm,height=5cm]{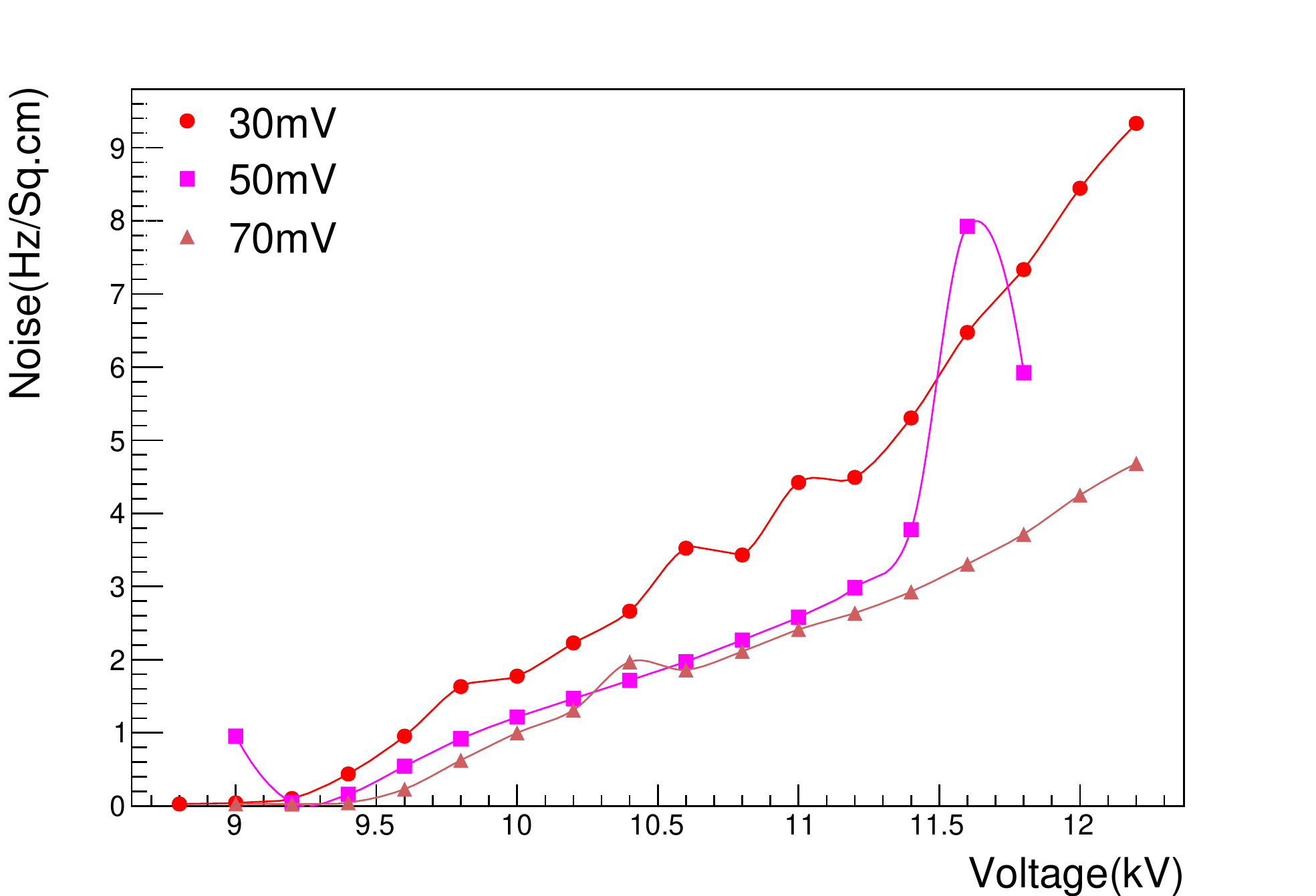}
             \end{minipage}
         \end{minipage}
          \caption{Efficiency and Noise Rate as a function of Threshold variation for $R134a$ (95.0\%), $C_{4}H_{10}$ (4.5\%), $SF_{6}$ (0.5\%) gas mixture.}
\label{fig:Threshold}
 \end{figure}

The discriminator threshold value set to reduce the systematic noise may 
also affect the performance of the detector by suppressing the signal. To study this we 
varied the threshold values for the RPC and scintillator paddles and measured the 
efficiencies and noise rates. Fig.~\ref{fig:Threshold} shows the effect of threshold variation on the efficiency and noise rates for Asahi glass RPC. We did not 
observe significant dependence in efficiency, whereas the noise rate decreases as we increase 
the threshold value from 30mV to 50mV. However, further increasing the threshold 
value to 70mV did not affect the noise rate much, so we fixed the threshold value 
of 50mV for operating the RPCs.

\section{Results and Conclusions}\label{sec:res}
The INO-ICAL experiment will be required to build approximately 28,000 RPC's of size $2~m \times 2~m$. Because of the requirement of large number of RPCs it is very important to perform a through $R\& D$ on all aspects of the detector performance before finalizing various 
parameters. We procured three types of glasses available in the local 
market, viz. Asahi, Saint Gobain and Modi to perform our $R\& D$. We performed various studies to assess 
the material and electrical properties of these glasses. We found Asahi to be best in terms of surface resistivity as well as smoothness. Saint Gobain is best in terms of Bulk Resistivity and transmittance. Asahi and Saint Gobain is comparable in terms of smoothness and transmittance whereas Modi is lacking in almost all the parameters.

From detector performance studies, we found that all the three makes gives almost similar 
efficiencies. However, Asahi RPC is having the minimum noise and leakage current 
with Saint Gobain comparable to Asahi under $R134a$ (95.0\%), $C_{4}H_{10}$ (4.5\%), $SF_{6}$ (0.5\%) gas composition. From gas composition variation studies we found that the efficiency is almost independent  upon the gas composition.  The behaviour is 
similar for all types of glasses. The gas composition of $R134a$ (95.0\%), $C_{4}H_{10}$ (4.5\%), $SF_{6}$ (0.5\%) was found to be having optimum efficiency and noise rate. The variation of temperature and 
humidity showed that with increase in temperature from $17^{\circ}C$ to $21^{\circ}C$ 
and relative humidity variation between approximately 40 and 60\%, the noise increases considerably. But, further increasing the temperature to $23^{\circ}$C did not affect the noise much. 
Leakage current also increases slightly with increase in temperature from $17^{\circ}C$ to 
$21^{\circ}C$ but further increase in temperature to $23^{\circ}C$ did not have any impact 
on current. We did not find any considerable effect of temperature and relative 
humidity on efficiency. On the 
other hand, we found that humidity has some weird effect on the current. Higher humidity
increased the currents considerably and at 67\% relative 
humidity we observe strange behavior in current which is shown in Fig.~\ref{fig:Temp} (centre). We are further investigating this unexpected 
behaviour.

From the discriminator threshold variations studies, we found that by increasing the threshold from 30mV 
to 50mV the systematic noise decreased considerably, but further increasing the 
threshold from 50mV to 70mV did not have any effect on noise rate. Therefore, we set the threshold value to 50mV for operating the RPCs.

In conclusion, we found that Asahi glass is best suited for the INO-ICAL RPC in terms of 
almost all the parameters that we studied. Saint Gobain, however, is also not far behind 
and appears to be equally good for INO-ICAL RPCs. The gas composition of $R134a$ (95.0\%), $C_{4}H_{10}$ (4.5\%), $SF_{6}$ (0.5\%) appears to be optimum. It is important to maintain the environment temperature around $21^{\circ}C$ and atmospheric relative humidity under control to operate INO-ICAL RPCs. All our results are based on small RPCs of size $30~cm \times 30~cm$, therefore, in order to complete our studies we are in the process of fabricating full size $2~m \times 2~m$ RPCs to confirm our measurements and findings.

\acknowledgments

We acknowledge the financial support from Department of Science and Technology (DST) and University of Delhi $R\&D$ grants for carrying out these studies. Daljeet Kaur would like to thank Council of 
Scientific and Industrial Research (CSIR) for the financial support. We would also like 
to thank the INO group at Tata Institute of Fundamental Research (TIFR) for providing 
some of the raw materials for detector construction. We also thank Prof. J. P. Singh of IIT Delhi for his help in getting AFM measurements done at IIT Delhi.


\begin{thebibliography}{9}
\bibitem{ino}INO Collaboration, India-based Neutrino Observatory Report, INO-2006-01.
\bibitem{Santonico1981}R. Santonico, R. Cardarelli, Nucl. Instr. and Meth. \textbf{A 187} (1981) 377. 
\bibitem{Santonico1988}R. Santonico, R. Cardarelli, Nucl. Instr. and Meth. \textbf{A 263} (1988) 20.
\bibitem{Bencze} G. Bencze et al., Nucl. Instr. and Meth. \textbf{A 340} (1994) 466.
\bibitem{ourbakelite} A. Kumar, Ankit Gaur, Md. Hasbuddin, Praveen Kumar, Purnendu Kumar, Daljeet Kaur, Swati Mishra and Md. Naimuddin, Submitted for publication in JINST.
\bibitem{Pramana} Sarika Bhide, V.M. Datar, Satyajit Jena, S.D. Kalmani, N.K. Mondal, G.K. Padmashree, B. Satyanarayana, R.R. Shinde, P. Verma, Pramana, Journal of Physics \textbf{69(6)} (2007) 1015.
\bibitem{gas}MA Lie-Hua, WANG Yi-Fang, Chinese Physics C \textbf{34} (2010) 1116.



\end{thebibliography}
\end{document}